\documentclass[referee,usenatbib]{mn2e}
\usepackage{epsfig}
\newcommand{\neuthy}{H{\sc i}}

\begin{document}
%Version 3.8
% Dealt with the corrections.

%      Comments 1,5 and 7 are basically the same, which if I can
%      paraphrase it as "Doesn't look like it too me". Well it does to
%      me - so where do we go from here? 

\title[On the association of G343.1-2.3 and PSR 1706-44]{On the
association of G343.1-2.3 and PSR 1706-44} 

\author[Dodson \& Golap]{R. Dodson $^1$,
  K. Golap $^2$ \\
$^1$ School of Mathematics and Physics, University of Tasmania, 
     GPO Box 252-21, Hobart, Tasmania 7001, Australia;\\  
     Richard.Dodson@utas.edu.au\\
$^2$ National Radio Astronomy Observatory, P.O. Box 0, Socorro, NM 87801\\
     kgolap@nrao.edu\\}

\maketitle

\begin{abstract}
The association of G343.1-2.3 and PSR 1706-44 has been controversial
from its first proposal. In this paper we address the difficulties,
and argue that the association is still likely. New evidence comes
from images of G343.1-2.3 obtained using the Australia Telescope
Compact Array (ATCA), and the pulsar obtained using the {\it CHANDRA}
X-ray observatory. Mosaicing was required to cover the full extent of
G343.1-2.3, and we present the polarisation images from this
experiment. Also an X-ray pulsar wind nebula has been found in the
archived {\it CHANDRA} observations, with the correct morphology to
support the association. The ATCA observations confirm the much
larger extent of the SNR, which now encompasses the pulsar. The X-ray
morphology points back toward the centre of the SNR, indicating the
direction of the proper motion, and that the PSR and SNR are
associated.
% max 200 words, currently about 150
\end{abstract}

% |HAM-2-2| _ HAMLET That you must teach me. But let me conjure you, by 
% |HAM-2-2| _ the rights of our fellowship, by the consonancy of 
% |HAM-2-2| _ our youth, by the obligation of our ever-preserved 
% |HAM-2-2| _ love, and by what more dear a better proposer could 
% |HAM-2-2| _ charge you withal, be even and direct with me, 
% |HAM-2-2| _ whether you were sent for, or no? 


\section{Introduction}

The pulsar PSR B1706-44 is one of only seven that are known to emit
gamma-rays in the GeV energy range \citep{thompson_92} and one of only
three that have been detected in the TeV range
\citep{kifune_95,chadwick_98}. \citet{ros_1706} reported that the
power law spectrum from the compact source at the site of the pulsar
was best explained by synchrotron origin, presumably a nebula around
the pulsar. \citet{finley_98} found that PSR B1706-44 is surrounded by
a compact X-ray nebula of radius $\sim0.3$ pc using {\it ASCA} data. This is
also true of the other two TeV emitters, the Crab and Vela
pulsars. The TeV gamma-rays, which appear to be unpulsed, can be
explained as resulting from the Inverse Compton interaction of the
high energy electrons, which produce the X-ray synchrotron emission,
with ambient photons \citep{aharonian_97}.

%     Further more in the Becker paper they do not see the PWN, but
%     deduced it from the spectrum. The ROSAT PSF is much greater than
%     the observed source size. I will make this clear in the revised text.

The original discovery that the {\it COS-B} gamma-ray source was a
pulsar was made by \citet{PSR1706_discover}. The spin down luminosity is
high, $3.4 \times 10^{36}$~ergs, as would be expected. \citet{1706}
published a map made by the MOST telescope at 843~MHz of the area
around the gamma-ray source. It showed a semicircular arc of emission,
which has subsequently been denoted as the supernova remnant (SNR)
G343.1-2.3, with the pulsar seemingly embedded in a small feature at
its south eastern extremity. It was argued that the approximate
distance of 3~kpc for the remnant, derived from the surface
brightness-diameter relationship ($\Sigma$-D) was compatible with the
pulsar's dispersion measure, although the distance indicated by the
widely used interstellar electron density model of \citet{TC_model}
would be only 1.8~kpc. The $\Sigma$-D relationship was based on the
flux density values from single dish observations, as both the VLA and
the MOST integrated flux fell short of what would be expected,
implying that the broad structure was being resolved \citep{1706}.

If the association of the pulsar and the supernova remnant is real, a
transverse velocity of $\sim900$~kms$^{-1}$ is required for the
pulsar to have moved from the approximate geometric centre of the SNR
arc in the characteristic spin down time of 17~kyr. This is high,
but not the highest measured value for a pulsar, nor as high as
that implied for some seemingly well-established pulsar-SNR
associations \citep{duck_1,duck_2}.

\citet{young_pulsars} imaged the area around the pulsar at 20~cm and
90~cm with the VLA and cast doubt on its association with
G343.1-2.3. These arguments were based on the morphology, and the fact
that the dispersion measure (DM) and $\Sigma$-D distance were
discrepant. The latter was subsequently countered by a 21~cm hydrogen
line absorption measurement of the pulsar by \citet{korbalski_HI}
which gives a distance range of 2.4 to 3.2~kpc. Most recently the
region has been imaged in more detail by \citet{giacani_01} at 20, 6
and 3.6~cm and they concurred with \citet{young_pulsars} description
of a nebula size of $3.5^\prime \times 2.5^\prime$.

\citet{johnston_1706} also argue against the association. They report
a maximum value of only 27~kms$^{-1}$ for the magnitude of the transverse
velocity of the pulsar calculated from the interstellar
scintillation. However in %both \citet{richard_phd} and then 
their later paper \citep{scint_49}, this estimate was revised up to
100~kms$^{-1}$. The possibility of a direct determination of the
proper motion of PSR B1706-44 has been investigated, but no phase
reference sufficiently strong for the current Australian VLBI network
(the LBA) could be found.

We have made a number of high sensitivity, low surface brightness
images of the remnant to determine its full extent. Furthermore we have
used {\it CHANDRA} archived data to search for a pulsar wind nebula
(PWN) to provide evidence for the pulsar interaction with the
interstellar medium (ISM).

\section{Results from the Compact Array}

The Australia Telescope Compact Array has been used to re-observe
G343.1-2.3, with a nineteen pointing mosaic. Mosaicing was required 
both to cover the extent of the source and also to assist in the recovery of
the flux on the shortest {\em uv} spacings, which were missed in the
MOST and the VLA observations.  Primary calibration was against
PKS1934-638 which has an integrated flux of 16.4~Jy at 1.384~GHz.
The mean observing epoch was September 1998. Deconvolution used the
mosaic version of the standard {\small MIRIAD} maximum entropy
methods for polarised images; {\small PMOSMEM}. The RMS
noise in the final total intensity image is 0.6~mJy.

\begin{figure}
 {
   \begin{center}
     \begin{minipage}[t]{0.45\textwidth}
         \psfig{file=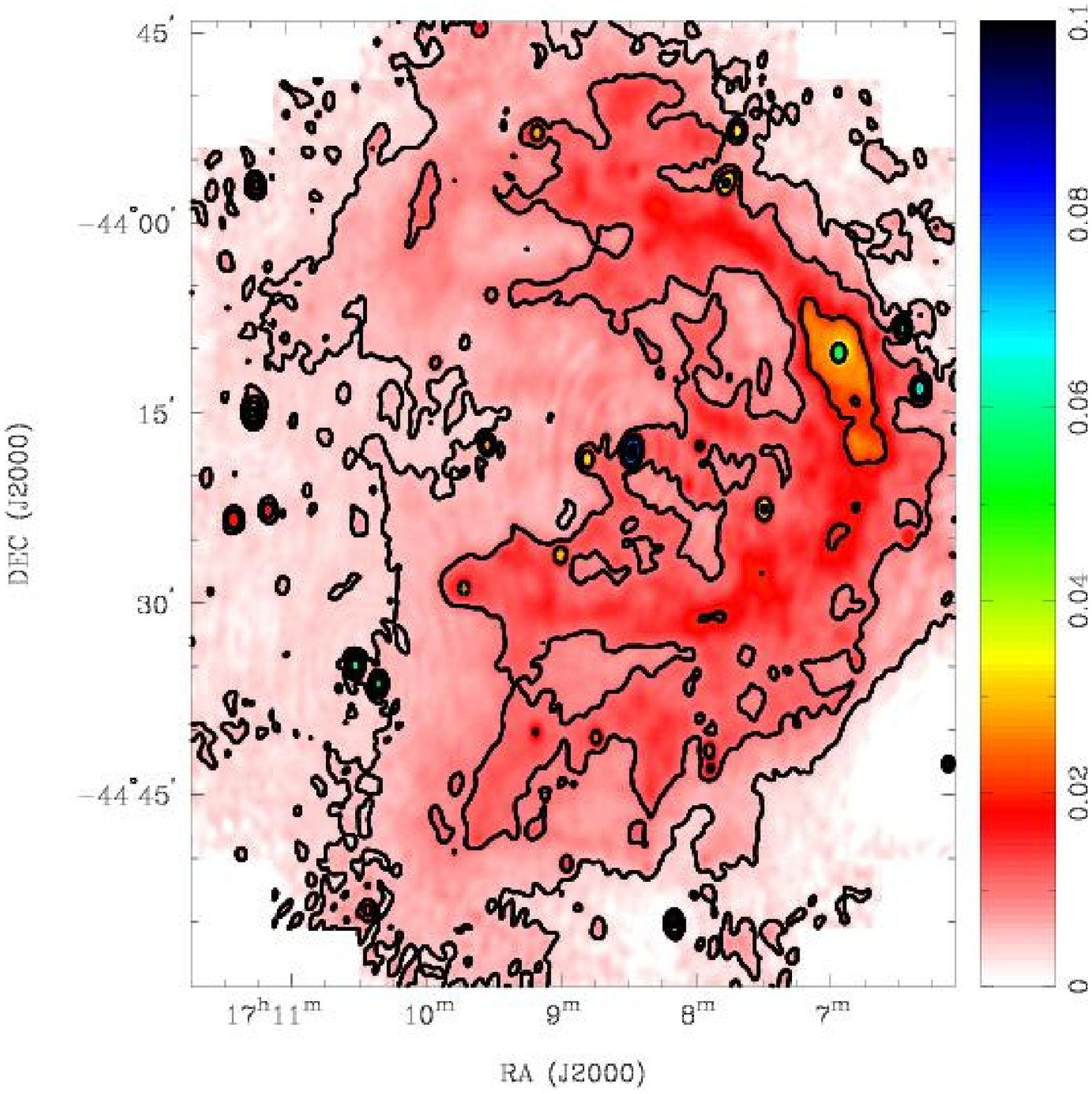,height=1.0\textwidth,angle=270}
     \end{minipage}
     \hfill
     \begin{minipage}[t]{0.45\textwidth}
         \psfig{file=pwn.eps,height=1.0\textwidth,angle=270}
     \end{minipage}
   \end{center}
 }
 \caption{Supernova Remnant G343.1-2.3 at 1384 MHz. a) the full
remnant and b) the region around the pulsar, with polarisation angles
overlaid; a 10~mJy polarisation scale bar is shown} 
 \label{fig:at_i}
 \end{figure}

Figure \ref{fig:at_i} shows the total intensity images of the whole
remnant (Figure \ref{fig:at_i} a) and of the region around the PSR
(Figure \ref{fig:at_i} b). The pulsar region is overlaid with the
polarisation vectors. The full image is very similar to those from
MOST \citep{1706}, and indeed has very similar resolution
($70^{\prime \prime} \times 47^{\prime \prime}$) (the longer
baselines, to the 6km antenna, were not used due to the {\em uv} poor
coverage). We have included as a constraint in the deconvolution the
observed total flux for the SNR, 30 Jy, observed with the University of
Tasmania's 26m antenna.

When the total flux from single dish measurements is included, the
images obtained are in agreement with the other observations, i.e. the
extent agrees with the very low frequency images from the VLA
\citep{young_pulsars} and MRT \citep{richard_phd,kumar_phd}, and the
spectral index has a simple monotonic spectrum of -0.5
%(though that is heavily influenced by the single
%dish information) 
\citep{iau199}.

\begin{figure}
\begin{center}
\epsfig{file=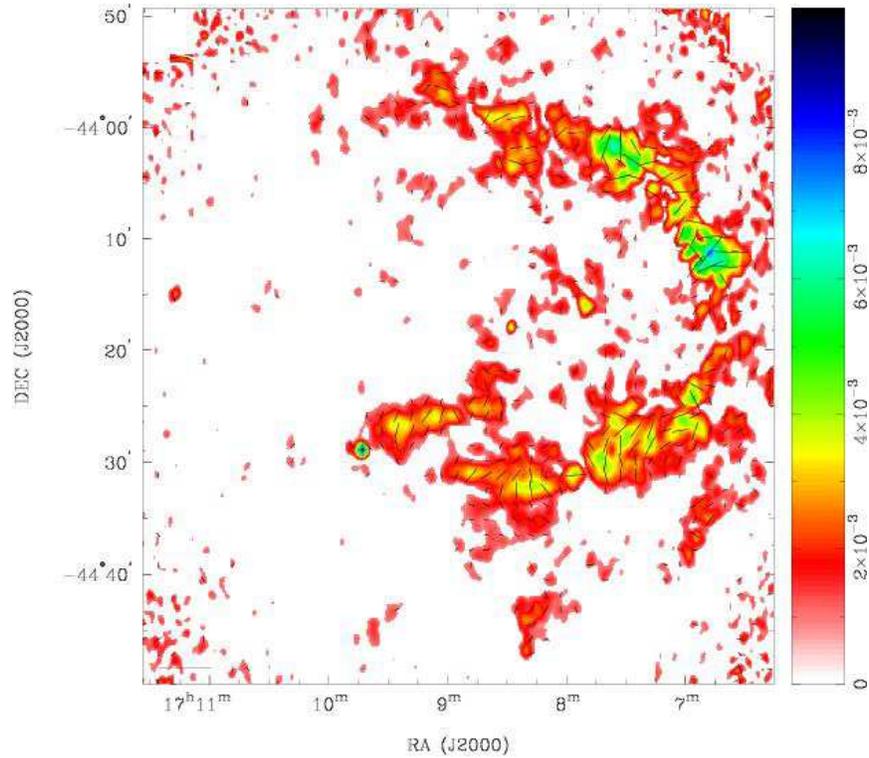,width=10cm,angle=270}
\caption{G323.1-2.3 polarised intensity, with polarisation angle
overlaid; a 10~mJy polarisation scale bar is shown}
\label{fig:at_pol}
\end{center}
\end{figure}

Figure \ref{fig:at_pol} shows the polarisation angle overlaid on the
polarised intensity. The polarised fraction is typically about 20
percent with a randomly orientated polarisation angle which changes
rapidly around the bright ring. Wisps of polarised emission can be
seen to extend beyond the  originally defined ring, and throughout
the region with broadscale emission. This confirms that the broadscale
emission is associated with the synchrotron emission from the SNR.

The pulsar is a point source with a flux density of 10~mJy, in line with that
found in the scintillation fluctuation observations of \citet{scint_49},
and is 70 per cent linearly polarised, and 25 per cent circularly
polarised. The mean pulsar position is 17:09:42.71 -44:29:08.1 with an
RMS of 0.5 arc seconds between epochs, within the errors estimates
of \citet{wang2000}, but is slightly south of the VLA observations;
that is -44:29:06.6.

It is instructive to compare the VLA images of \citet[figure
4]{young_pulsars} with the image in Figure \ref{fig:at_i} b). The
pulsar lies on a spur of emission which is clearly a continuous
feature covering more than $30'$. Thus it is extremely unlikely to
be a cometary tail from the pulsar. Furthermore the width of the spur,
$4'$, is too broad to be due to ram pressure of a moving
pulsar. It is intriguing that this is approximately the size that would be
expected from a static pulsar wind, derived from the Sedov equations with a
continuous input of energy, and with the $\dot{E}$ of PSR1706-44
\citep{pwn_first}. Taking the expressions from \citet{PWN_survey}, the
equation of the static shock radius ($R_s$) is;

% Edot = 10^(36.533496737429)
% D_kpc  2.4 to 3.2~kpc.
%age 10^4.2417017311165
\begin{equation}
R_s = ( \frac{\dot{E}}{4 \pi \rho_0} )^{1/5} t^{3/5}  {\rm pc} \approx \\
3.6 \dot{E}_{34}^{1/5}t_5^{3/5}n_0^{-1/5} D_{kpc}^{-1} {\rm \ arc\ min}
\label{eq:pssn}
\end{equation}

Where $\dot{E}_{34}$ is the spin down energy in unit of
$10^{43}$~ergs~s$^{-1}$, $n_0$ is the number density, $t_5$ is the age
in $10^5$ years and $D_{kpc}$ is the distance in kpc. Using the
standard values gives $1.7^{\prime} n_0^{-1/5}$ using the lower limit
on the distance. 

The formula for a ram pressure shock driven PWN ($R_w$) is;

\begin{equation}
R_w = ( \frac{\dot{E}}{4 \pi c\rho \nu^2_{PSR}})^{1/2} \approx 
0.475 \dot{E}_{34}^{1/2} n_0^{-1/2}\nu^{-1}_{150} D_{kpc}^{-1} 
{\rm \ arc\ sec}
% R_w = 475 \dot{E}_{34}^{1/2} n_0^{-1/2}\nu^{-1}_{150} AU
% AU = 2.06E-5 PC
\label{eq:pwn}
\end{equation}

or $3.7^{\prime \prime} n_0^{-1/2}\nu^{-1}_{150}$. $\nu_{150}$ is the
velocity in units of 150 kms$^{-1}$.

\vspace{1mm}

It is clear that the radius of the feature seen in the radio band cannot be
due to the ram pressure shock unless the density is extremely low ($2
\times 10^{-4}$). It is consistent with the size expected for a
static PWN, with a number density of the order suggested by the
analysis of the {\it ROSAT} data, i.e. 0.09~cm$^{-3}$
\citep{richard_phd,iau199,ros_1706}. These observations are
contaminated by the nearby low mass X-ray binary, LMXB~1705-44, but a
best fit of a thermal spectrum to the X-ray emission from the SNR
gives a distance of 3.1~kpc, and an age of 8.9~kyrs. These values are
derived using the X-ray luminosity to distance model of
\citet{rosat_snr_b}. The local density based on the pulsar age is
0.09~cm$^{-3}$, whilst that from the Sedov model age is
0.02~cm$^{-3}$.

%      The second and third points refer to the paper of 1985 with
%      Becker, Braizer & Trumpter. This is referenced, where I thought
%      it was applicable. I downloaded the ROSAT data from the public
%      archives used in my thesis (Dodson '97) and a conference paper
%      (Dodson et al '99). In this work I managed to remove the
%      confusing source to reach several conclusions about the
%      SNR. However I have now referenced their paper.

\section{Observations with the {\it CHANDRA} X-ray telescope}

We took 48 ksecs of {\it HRC} data and 17 ksecs of {\it ACIS} data
from the {\it CHANDRA} X-ray observatory archive and imaged the region
around the pulsar. These observations were made in August and February
2000, with roll angles 180$^o$ apart. A similar low surface brightness
extended feature is clearly visible in both the {\it HRC} and the {\it
ACIS} data, as first reported by \citet{finley_98}. Their {\it ASCA}
observations, complicated by contamination from the bright
LMXB~1705-44, saw evidence for a 14--33$^{\prime \prime}$ compact
nebula in the radial surface brightness. The feature as seen by {\it
CHANDRA} as it is visible in both data sets, can not be due to the
space craft dither, nor due to a soft proton flare or any other random
and unrelated event.

%      Comment 4; I used the standard radio astronomical deconvolution
%      method, CLEAN. I can expand on this method if needed, or would it
%      be sufficent to make it clear that it is this technique I am
%      using? 

% 4: In Chapter 3: 'we have found the peak energy of the pulsar 
% emission to be around 1 keV from ACIS data, and have 'cleaned' 
% the HRC data with the PSF for that energy'. This does not make 
% clear what the authors really did in their analysis. It should be 
% described in much more detail. If the authors speak of diffuse 
% emission, they should address the question if the Chandra 
% observation was affected by soft-proton flares which eventually 
% could mimic this diffuse emission. The Chandra satellite is 
% wobbling (does a jitter). The authors should discuss what the 
% wobbling direction is in their observation. They further should make 
% clear that the 'orientation' of the diffuse feature they find is not due 
% to incomplete jitter correction.  

\begin{figure}
\begin{center}
\epsfig{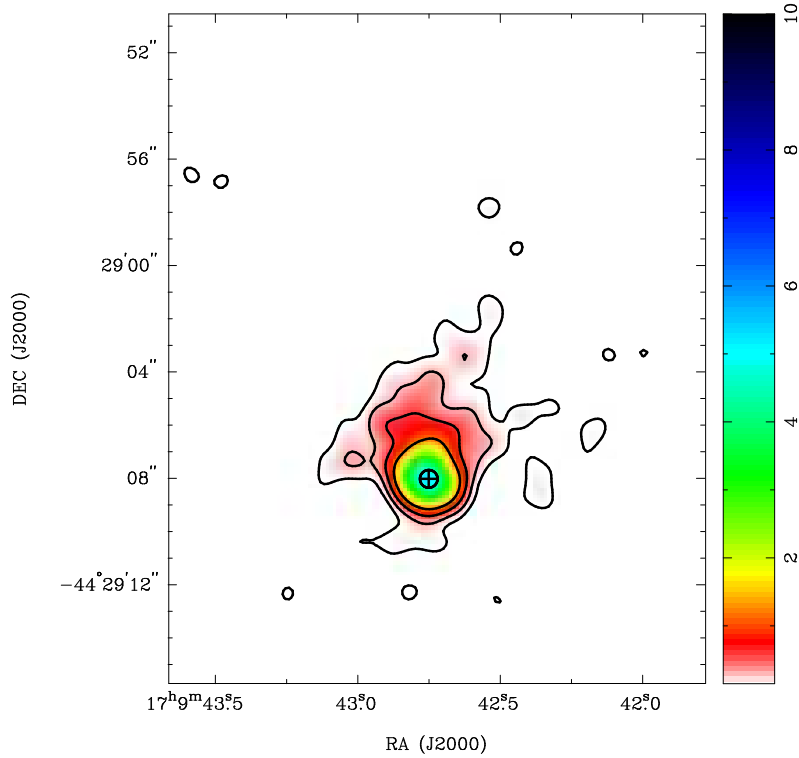}
\caption{PSR1706-44 PWN in X-rays, from the Chandra HRC. Scale is in
counts per dector pixel, the 5, 10 and 20 $\sigma$ contours are
shown. The FWHM of the used HRC PSF is shown at the site of the
pulsar.} 
\label{fig:chandra}
\end{center}
\end{figure}

We have found the peak energy of the pulsar emission to be around 1
keV from the {\it ACIS} data, and have `CLEANed' the {\it HRC} data
(smoothed to 0.3$^{\prime \prime}$) with the PSF for that energy.
CLEAN is a standard deconvolution technique where a small fraction of
the PSF is subtracted interatively from the image, until a conversion
condition is met \citep{hogbom_74}. Our convergence criterion was
satisfied when the peak emission was more than half the PSF width from
the pulsar position. Significant emission (10 percent of peak)
remained at this point. It was perfectly feasible to clean this
feature iteratively as well, but equally good results were obtained by
fitting a Gaussian model around the pulsar position. Figure
\ref{fig:chandra} shows the emission after the cleaning the point
source with the PSF; its position is marked with a cross. This
confirms the object seen by \citet{finley_98}, but finds a two
dimensional structure and a smaller extent which can be explained by
the degree of contamination in the ASCA image. The object has a very
compact nebula (but much greater than the PSF) around the pulsar which
is best fitted by a Gaussian of about 1.1$\pm 0.1^{\prime \prime}$,
and a more tenuous trail pointing nor-north west, and extending about
5$^{\prime \prime}$ back from the pulsar. The angle of symmetry of
this tail is $-15^o$, toward the SNR centre. The brightness of the
tenuous nebula is about 10 times greater than that of the point spread
function of the {\it CHANDRA} mirrors; i.e. 1 percent of the pulsar
peak, whereas the PSF would be 0.1 percent. The most likely, and most
natural, physical model of such a structure is a bow shock, with the
more tenuous tail left behind in the wake.

% 1: The shape of the pulsar-wind nebula is not conclusive in respect 
% to the pulsar motion/direction. Unless the nebula is really identified 
% as a bow-shock or an X-ray tail behind the pulsar along its proper 
% motion direction. The small feature seen in the Chandra data can 
% not be claimed 'X-ray tail' and is not convincing in its interpretation 
% that it points to wards the (putative) pulsar birth place.  

% 5: The authors speak from a bow-shock in the Chandra data. But 
% the image based on the Chandra data does not show a bow-
% shock? Can they clarify what they mean?  

% 7: The conclusion that the pulsar is associated with the SNR as 
% drawn in Concl., is by no means convincing. The results obtained 
% from the Chandra data do not add to this argumentation at all! The 
% authors are therefore asked to rewrite the paper, making 
% statements which corresponds to the significance of their results. I 
% doubt that their results and ATCA images do straighten the 
% reasons for an association of PSR 1706-44 and G343.1-2.3  

The pulsar to bow shock distance at the closest approach ($1.1^{\prime
\prime}$), at 2.4 kpc, is just 0.05 pc. If the scintillation velocity
limit found by \citet{scint_49} is correct then the density is
required to be greater than 25~cm$^{-3}$, or, if the densities are
those found from the {\em ROSAT} observations
\citep{richard_phd,iau199,ros_1706}, the velocities are in the range
1000--2000~kms$^{-1}$, greater than that required to move from the
centre of the SNR to the current position. More typical densities
imply reasonable velocities (500 kms$^{-1}$ implies a density of
unity). These are obviously much higher that those seen by the
scintillation velocity, but we stress that this
is an indirect measurement with a large intrinsic scatter, and only a
VLBI observation will settle the question, for which the LBA will need
to be upgraded.

\begin{figure}
\begin{center}
\epsfig{file=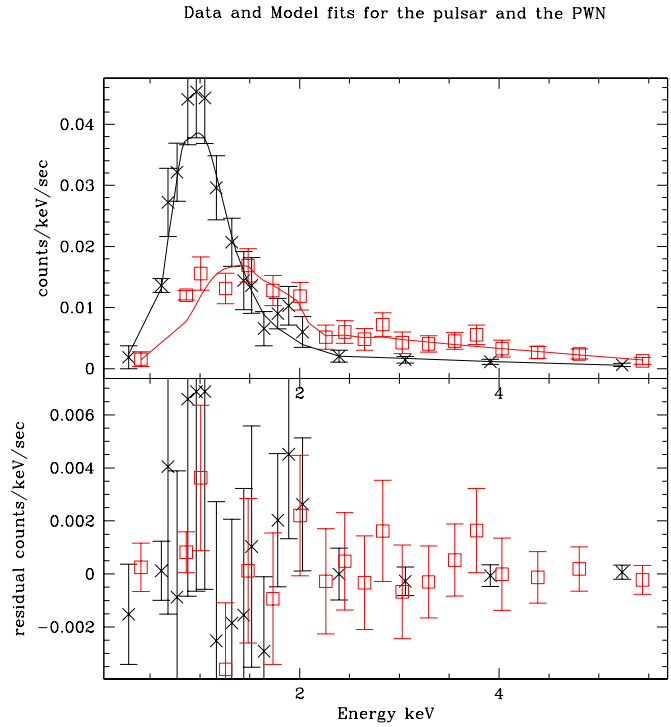,width=10cm}
\caption{Spectral fit to the pulsar region (black crosses) and the PWN
(red squares).} 
\label{fig:x-ray}
\end{center}
\end{figure}

From the {\it HRC} image we cleaned out 294 counts for the pulsar
before the convergence criterion was reached. The integrated X-ray
flux from the compact nebula is 390 counts and the tenuous tail is 120
counts. We fitted a number of models to the {\it ACIS} data in {\small
SHERPA 2.2}. The best fit to the source spectrum was found with joint
fitting of the two sources; the region around the pulsar as a black
body plus a power law portion and the PWN as an independent power
law. These were attenuated with the usual column density
absorption. The fact that the same spectrum could not be fitted to
both features confirms that it is not the result of incorrectly
removed spacecraft dither. The model fits for the black body can be
used to calculate the diameter of the emission region from the
normalisation constant ($228.5_{-138}^{+525} R_{km}^2
D_{10kpc}^{-2}$). Using the \neuthy\ distance limits gives the radius
of the emitting region as $3.6^{+3}_{-1.3}$--$4.8^{+4}_{-1.8}$~km. The
unabsorbed X-ray flux gives the range of intrinsic luminosities to be
%luminocity =  4*pi*[2.8-3.2]e3*3.1e18*flux=[5.8 8.1]e-13
5.4--7.1~$\times 10^{32}$~ergs~s$^{-1}$ from the black body
(presumably the pulsar) and 7.6--9.9~$\times 10^{32}$~ergs~s$^{-1}$
from the power law regions (the PWN). The fits are shown in figure
\ref{fig:x-ray}. The reduced Chi-squared is 0.6 and a Q-statistic of
99\%, implying that the errors have been underestimated. The column
densities agree with those from the {\it ROSAT} observations
\citep{ros_1706}.

%      comment 6, on the spectral models. "it is hard to recognize
%      something in Fig.4. The plot should be strongly improved" What is
%      hard to recognise? How would he like it improved? I will sum up
%      the number of points, as if it in black and white it is very
%      crowded. Why are the "standard bands" standard? I am happy to
%      comply, but as I said I'm a radio astronomer, not an X-ray
%      astronomer.

% 6: The different spectral models used to fit the Chandra data, as 
% well as the fits, should be discussed in more detail, with a table 
% and the reduced Chi-Sqr numbers (and goodness of fits) listed. In 
% respect to this, it is hard to recognize something in Fig.4. The plot 
% should be strongly (!) improved. The X-ray flux should be given for a 
% standard energy band (0.1-2.4 keV and 0.5-10 keV or 2-10 keV 
% etc.) It should be discussed what parameters were fixed in the 
% spectral fits and which parameters were fitted.  

The fitted model parameters are;

\begin{tabular}{l l l l l}
$n_H/10^{21}$  & T$_{BB}$/keV & $\gamma_{psr}$& $\gamma_{pwn}$&reduced $\chi^2$\\
\hline
$5.3\pm0.8$  & $0.15 \pm 0.02$ & $1.7 \pm 0.5$ & $1.5 \pm 0.2$&31/53\\
\hline
\end{tabular}

The integrated X-ray flux between 0.5 and 10 keV for each component
in ergs~cm$^{-2}$s$^{-1}$ is;

\begin{tabular}{l | l l l}
	   & Black Body        & Power Law$_{psr}$ &  Power Law$_{pwn}$ \\
\hline
Received   &$0.7\times10^{-13}$ &$1.6\times10^{-13}$ &$4.6\times10^{-13}$ \\
Unabsorbed &$5.8\times10^{-13}$ &$2.2\times10^{-13}$ &$5.9\times10^{-13}$ \\
\end{tabular}

%The received and unabsorbed total X-ray flux across the fitted data
%set (0.3 to 6 keV) is 2.3 $\times 10^{-13}$~ergs~cm$^{-2}$s$^{-1}$ and
%1.3 $\times 10^{-12}$~ergs~cm$^{-2}$s$^{-1}$, xxx $\times
%10^{-13}$~ergs~cm$^{-2}$s$^{-1}$ and xxx $\times
%10^{-12}$~ergs~cm$^{-2}$s$^{-1}$ in the 0.5 to 10 keV energy band.

\section{Conclusion}

Initially the arguments against the association were based on Kaspi's
criteria \citep{kaspi_assoc}. These can be summarised in order of ease
of observability as; same position, same distance, same age, velocity
reasonable and proper motion correct. It was felt that the pulsar DM
distance and the SNR $\Sigma$-D distance did not match, and the pulsar
was at the edge of the SNR and had too low a velocity to have
travelled from the geometric centre of the SNR. Also there was no
morphological evidence to imply a velocity direction, and the
scintillation velocity was too low. Since then work
\citep{korbalski_HI} has shown that the DM distance was incorrect, and
our ATCA observations have found that the extent of the SNR has been
underestimated. The Chandra data shows a morphological signature best
interpreted as the path of the pulsar. The proper motion of the pulsar
is the remaining question, and work is under way to attempt to answer
that.

It is clear from both the low total flux values found by the MOST and
VLA maps that there is a smooth broad component to this SNR. The ATCA
observations confirm the extent of the broad emission, and that this
is non-thermal in character. Taking this smooth component as the true
limits of the SNR enlarges the extent and moves it southwards. This
places the pulsar PSR1706-44 within the SNR shell, rather than on the
rim, thus satisfying Kaspi's first criteria. 

The X-ray PWN observed by {\it CHANDRA} is consistent with the pulsar
travelling from the general direction of the SNR centre, with either a
very slow velocity (which would agree with the scintillation results) or
a low local density (in agreement with the {\it ROSAT} results).

We expect a SNR to be associated with PSR1706-44 because of its
youth. The association with G343.1-2.3 was rejected in the past, but
we have shown that the grounds for the rejection are not
compelling. The question of the pulsar velocity could be resolved by
finding a phase reference for VLBI observation, with either the SKA or
an upgrade to the current LBA. We will be able to rule out the upper
velocity limits with the ATCA within the next few years but will not
be able to measure the lower limits for a few decades.

Finally the result stresses the importance of improving the ROSAT
observations of this SNR with a mosaic over the whole area of the
radio emission using either {\it CHANDRA} or {\it XMM}.

\section{acknowledgments}

During the writing of this paper K.R. Anantharamaiah of the RRI passed
away. Both of us trained as astronomers there, and wish to acknowledge
his friendly encouragement and interest that persuaded us that radio
astronomy is a rewarding and exciting field.

Also we wish to thank B. Gaensler for discussion of the spectral
fits. 

The X-ray data for the Chandra observations were downloaded from the
public access site, the Chandra Data Archive (CDA). This is part of
the Chandra X-Ray Observatory Science Center (CXC) which is operated
for NASA by the Smithsonian Astrophysical Observatory. The Australia
Telescope Compact Array, funded by the Commonwealth of Australia for
operation as a National Facility, is managed by CSIRO. This research
made use of NASA's Astrophysical Data System Abstract Service.

%\bibliography{pap1706}
%\bibliographystyle{mn2e}

\onecolumn

\end{document}